\documentclass[aps,12pt]{revtex4}
\usepackage{amssymb}
\usepackage{graphicx,setspace}
\usepackage{amsmath}
\usepackage{mathrsfs}

\def\v{{\mathbf v}}

\def\f{{\mathbf f}}

\def\U{{\mathbf U}}

\def\bnabla{\mbox{\boldmath$\nabla$}}

\begin{document}
\title{The Basics of Water Waves Theory for Analogue Gravity}
\author{Germain Rousseaux}
\affiliation{Universit\'{e} de Nice-Sophia Antipolis,\\
Laboratoire J.-A. Dieudonn\'{e}, UMR CNRS-UNS 6621,\\
Parc Valrose,\\
06108 Nice Cedex 02,\\ France, European Union.}


\begin{abstract}
This chapter gives an introduction to the connection between the physics of water waves and analogue gravity. Only a basic knowledge of fluid mechanics is assumed as a prerequisite.

\end{abstract}

\maketitle

\begin{flushright}
  \begin{minipage}[h]{15cm}
    \begin{quote}
    \tiny
``How dare you roil my drink ? Your impudence I shall chastise !"\\
``Let not your majesty," the lamb replies, \\
``Decide in haste or passion !"  \\
``For sure It's difficult to think In what respect or fashion my drinking 
here could roil your drink, since on the stream your majesty 
now faces I'm lower down, full twenty paces".\\
      \begin{flushright}
Jean de la Fontaine
      \end{flushright}
    \end{quote}
  \end{minipage}
\end{flushright}

\begin{figure}[!htbp]
\includegraphics[width=6cm]{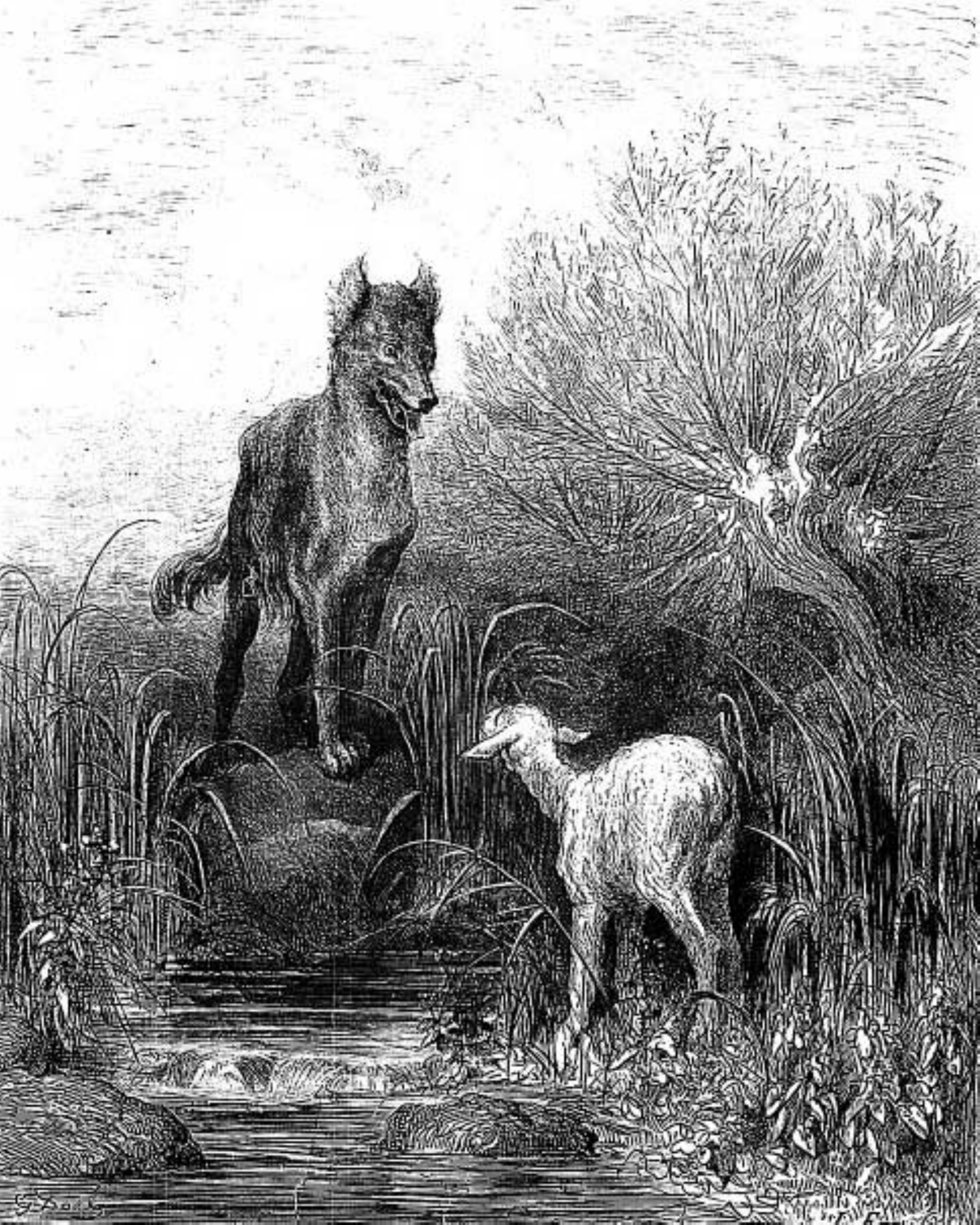}
\caption{The Wolf and the Lamb by Gustave Dor\'e.}
\label{WL}
\end{figure}

\newpage

\section*{Introduction}

According to Pierre-Gilles de Gennes, {\it ``the borders between great empires are often populated by the most interesting groups"}. Indeed, these people often speak several languages and are more open-minded due to cultural exchanges. A wonderful analogy exists between the propagation of hydrodynamic waves on a fluid flow and the propagation of light in the curved space-time of a black hole. It allows us to test astrophysical predictions such as Hawking radiation and the effects of high frequency dispersion on it. It provides new insights in Fluid Mechanics thanks to the use of tools and concepts borrowed from Quantum Field Theory in curved space-time and vice versa. General relativists speak with hydraulicians and this chapter is a testimony of their common language and relationships \cite{Ralph, BLV}.

Here, we provide the general background on water waves propagation for analogue gravity: we will try to explain how water waves propagate and how a flow current implies the existence of an effective space-time; we will insist on the difference between propagation in deep and shallow waters on the dispersion relation; a generalized definition of a horizon will be given and which, in the particular case of shallow water, reduces to the usual habit of general relativists.

\section{A glimpse of dimensional analysis}

The equations of fluid mechanics are known since several centuries but they still defy modern physics when we try to understand one of its outstanding mysteries like turbulence. As they are non-linear and feature several effects such as pressure, gravity and external forces, practitioners have been forced to introduce a very useful way to grasp the relevant effects when dealing with a peculiar flow. This technique is the so-called dimensional analysis which is often the only rescue procedure to disentangle the relative magnitude of several processes at play. The reader will be referred to the book by Barenblatt on scaling and dimensional analysis for a thorough introduction \cite{Barenblatt}. Here, we will construct with simple arguments the relevant velocities of propagation of water waves depending on the water depth. Three regimes of propagation will be uncovered with corresponding dispersion relations. Then, the effect of a current will be added and this ingredient is going to be essential in order to have an analogue gravity system.

Waves are characterized by both their wavenumber $k$ and their angular frequency $\omega$. Since we are dealing with Newton's laws of motion applied to fluids, the second time derivative (namely inertia) will translate in a square term in the angular frequency and the dispersion relation has the general form $\omega ^2 = F(k)$. Obviously, waves with both positive and negative angular frequency are thus described by the dispersion relation ($\omega =\pm \sqrt{F(k)}$). Usually, the negative root is dismissed since, when the propagation is free, it is a matter of convention to focus either on the right or left-propagating waves. Of course, in the presence of a current, the system will no longer be symmetric with respect to space reflection ($x \to -x$) and then both types of waves (positive and negative) turn to be important...

\subsection{Shallow waters}

Let us assume that a train of continuous sinusoidal water waves with wavelength $\lambda$ is propagating at the surface of a fluid at rest in a given depth $h$. We make the strong hypothesis that the wavelength is much longer than the depth, that is $kh<<1$ using the wavenumber $k=2\pi /\lambda$. The fluid vertical extension $h$ has the dimension of a length $L$. Since inertia is balanced by gravity, the gravity field is a relevant parameter and its intensity $g$ has the dimension of an acceleration $L.T^{-2}$. We are looking for the typical scaling of the wave velocity. The crests of the water waves propagate with the so-called phase velocity and its value $c$ has the following dimension $L.T^{-1}$. We are lead to the obvious scaling law up to a constant term :
\begin{equation}
c_{shallow} \approx \sqrt{gh}
\end{equation}

Using the definition of the phase velocity $c_\phi =\omega /k$, it is straightforward to infer the approximate dispersion relation for water waves propagating in shallow waters :
\begin{equation}
\omega ^2 \approx ghk^2
\end{equation}
which is similar to the dispersion relation for light waves in empty flat space-time. Analogue Gravity will emerge when effective curved space-time is added as we will see...

Gravity waves in shallow waters are not dispersive since, whatever their wavelength, they do propagate with the same velocity.

\subsection{Deep waters}

Far from the sea shore or for very short gravity waves, the water depth is no more a relevant parameter and the only length scale left is the wavelength of the water waves. So, if we assume that $kh>>1$ and recalling that the wavenumber $k$ has the dimension of an inverse length $L^{-1}$, we find the scaling for the phase velocity in deep waters:
\begin{equation}
c_{deep} \approx \sqrt{\frac{g}{k}}
\end{equation}
and the dispersion relation:
\begin{equation}
\omega ^2 \approx gk
\end{equation}

Newton derived this scaling in his Principia by applying the Galileo formula for the period of oscillation of a pendulum $T\simeq \sqrt{l/g}$ to the water waves. Indeed, if the length of the pendulum $l$ is replaced by the wavelength, we do recover the same scaling.

Gravity waves in deep waters are dispersive since longer waves propagates faster than the short ones.

\subsection{Arbitrary water depth}

Without doing more calculations, we can anticipate that the general dispersion relation for waters on arbitrary depth will write according to the following form :
\begin{equation}
\omega ^2=gk*H(k)
\end{equation}
bearing in mind that the following asymptotic limits must be fulfilled : $\lim_{kh \to \infty}H(k)= 1$ and $\lim _{kh \to 0}H(k)=kh$. It can been shown that one has rigorously $H(k)=\tanh{(kh)}$ after a lengthy calculation that we will avoid to the reader (see the books by Mei \cite{Mei} or Dingemans \cite{Dingemans} for a mathematical demonstration).

\subsection{The capillary length}

Water is made of molecules. In the bulk of the liquid, every molecule is surrounded by the same number of neighboring molecules whereas at the boundary with another substance (like air at the free surface of sea water), a water molecule is also surrounded by gas molecules and the resulting imbalance in the chemical interactions results in a pressure difference between the liquid and the gas. The Young-Laplace law states that the pressure jump is proportional both to the local curvature of the interface and to a phenomenological coefficient $\gamma$ named surface tension which is a property of both media :
\begin{equation}
\Delta p_{Y-L} = \gamma C
\end{equation}
where $C$ is the curvature of the free surface and has the dimension of an inverse length (roughly the radius of curvature).

When the Young-Laplace pressure $\Delta p_{Y-L}$ is balanced by the Pascal pressure drop $\Delta p_{P}=\rho g h$ due to the static gravity field, a scaling law is  deduced easily for the so-called capillary length which is the typical size on which surface tension effects are acting :
\begin{equation}
\Delta p_{Y-L} \approx \frac{\gamma}{l_c} =\Delta p_{P} \approx \rho g l_c
\end{equation}
Hence, the capillary length writes:
\begin{equation}
l_c = \sqrt{ \frac{\gamma}{\rho g}}
\end{equation}
Let us introduce the effective gravity field $g^*$ induced by the capillarity and its dispersive scaling law in terms of the wavenumber $k$:
\begin{equation}
g^* \simeq \frac{\gamma}{\rho} \frac{1}{l_c ^2} \approx  \frac{\gamma}{\rho} k^2
\end{equation}
The dispersion relation for water waves taking into account the effect of surface tension becomes:
\begin{equation}
\omega ^2=\left(g+g^*\right)k\tanh{(kh)}
\end{equation}
that is \cite{Mei}:
\begin{equation}
\omega ^2=\left(gk+\frac{\gamma}{\rho}k^3\right)\tanh{(kh)}
\end{equation}

\newpage

\section{Long water waves on a current as a gravity analogue}

In a letter to H. Cavendish in 1783, Reverend John Michell introduced the concept of what is named, in modern physics, a ``black hole" (he was inspired by the corpuscular theory of light by I. Newton) \cite{Schaffer}: {\it ``If the semi-diameter of a sphere of the same density as the Sun in the proportion of five hundred to one, and by supposing light to be attracted by the same force in proportion to its [mass] with other bodies, all light emitted from such a body would be made to return towards it, by its own proper gravity"}. According to him, this situation occurs when the ``escape velocity" of a massive particle is equal to the velocity of light. Then, Pierre-Simon de Laplace introduced the term ``{\it \'etoile sombre}" (dark star) in his ``Exposition du Syst\`eme du Monde" in 1796 to denote such an object. Laplace is also well known for having proposed in 1775 an analytical model to describe standing water waves in shallow water and for having derived the related dispersion relation \cite{Darrigol}. 

Recently, Sch\"utzhold \& Unruh derived the equation of propagation of water waves moving on a background flow in the shallow waters limit \cite{SU} and this equation describes also the behavior of light near the event horizon of a black hole. Indeed, under the impulsion of the seminal work by Unruh \cite{Unruh}, there has been more and more interest for analogue models in general relativity in order to understand the physics of wave propagation on an effective curved space-time. Several systems exhibit a so-called ``acoustic" metric similar to the metric describing a black hole when a wave is moving in a ``flowing" medium \cite{Ralph, BLV}.

Here, we will reproduce the derivation of Sch\"utzhold \& Unruh with some details for pedagogy \cite{SU}. Hence, we consider the propagation of small linear perturbations of a free surface between water and air in the presence of an underlying current. The current is uniform in depth $z$, time-independent and varies slowly in the longitudinal direction $x$. We are in the so-called WKBJ approximation such that the wavelength is smaller than the typical length on which the current varies ($\lambda <<\frac{U(x)}{\frac{dU}{dx}}$). $\U=\v_B$ will denote the background flow current whereas $\v$ will stand for the velocity associated to the propagation of waves.

The liquid is inviscid, its density is constant ($\rho=\mathrm{const}$) and the flow is incompressible.

We have the following equations of motion :\\
- the continuity equation which comes from the incompressibility condition: $\bnabla.\v=0$ \\
- the Euler equation : 
\vspace{-0.2cm}
\begin{equation}
\frac{d\v}{dt}=\dot{\v}+(\v.\bnabla)\v=-\frac{\bnabla p}{\rho}+ {\bf g} +\frac{\f}{\rho}
\end{equation}
with $p$ the pressure, ${\bf g}=-g\vec{e_z}$ the gravitational acceleration and $\f=-\rho \bnabla_\parallel V^\parallel$ a horizontal and irrotational force in the $x$ direction ($\parallel$) driven by the potential $V^\parallel$ which is at the origin of the flow.\\

Since the flow is assumed to be vorticity-free $\bnabla \times \v =0$ (a crucial feature of water waves propagation), one has $(\v.\bnabla)\v=(\bnabla \times \v )\times \v+\frac{1}{2}\bnabla(v^2)=\frac{1}{2}\bnabla(v^2)$ with $\v=\bnabla\phi$ where $\phi$ stands for the velocity potential. The vectorial Euler equation reduces to the simpler scalar Bernoulli equation \vspace{-0.2cm}
\begin{equation}
\dot{\phi}+\frac{1}{2}(\bnabla\phi)^2=-\frac{p}{\rho}-gz-V^\parallel
\label{E2}
\end{equation}

The boundary conditions are such that:\\
- in $z=0$, the vertical flow velocity must be null, i.e. $v^\perp(z=0)=0$. $z=0$ is by definition the bottom depth.\\
- the height variations of the fluid are determined by the very same velocity but computed on the free surface:
\vspace{-0.2cm}
\begin{equation}
v^\perp(z=h)=\frac{dh}{dt}=\dot{h}+(\v.\bnabla)h
\label{condition2}
\end{equation}
where $\frac{dh}{dt}$ is the velocity of a point on the air-water interface.\\
- the relative pressure with respect to the atmospheric pressure on the free surface cancels by definition : $p(z=h)=0$.\\

Let us consider a velocity perturbation $\delta v$ of the background flow $\v_B$ with a corresponding vertical displacement $\delta h$. We assume the background flow $\v_B$ to be stationary, irrotational and horizontal: $\bnabla_\perp \v_B=0$, $\v_B=\v^\parallel_B \rightarrow \bnabla_\parallel .\v_B=0$.\\

The Bernoulli equation gives:
\vspace{-0.2cm}
\begin{equation}
\frac{1}{2} v_B^2=-\frac{p_B}{\rho}-gz-V^\parallel
\label{eulervb}
\end{equation}
where $p_B$ follows Pascal's law for static pressure distribution in the water column $p_B(z)=\rho g(h-z)$.\\

We assume that the velocity perturbation $\delta v$ is also curl-less: hence, we can define a perturbed velocity potential $\delta \phi$. Using again the Bernoulli equation, we get:\\
\vspace{-0.2cm}
\begin{equation}
\delta\dot{\phi}+\v^\parallel_B .\bnabla_\parallel \delta \phi=-\frac{\delta p}{\rho}
\label{bernoullipert}
\end{equation}

By taking into account the condition $p_B(z=h)=0$ and using the expression for $p_B$, we obtain the boundary condition for the pressure fluctuation at the free surface $\delta p$: $\delta p(z=h)=g\rho \delta h$.

The same procedure applies to the vertical velocity, using the following conditions (\ref{condition2}) and $v^\perp(z=0)=0$:
\begin{tabular}{ccc}
$\delta v^\perp(z=0)=0$ & and & $\delta v^\perp(z=h_B)=\delta \dot{h}+(\v^\parallel_B .\bnabla_\parallel)\delta h$
\end{tabular}.

We now develop the velocity potential $\delta\phi$ using a Taylor series:
\vspace{-0.2cm}
\begin{equation}
\delta\phi(x,y,z)=\sum\limits_{n=0}^{\infty}\frac{z^n}{n!}\delta\phi_{(n)}(x,y)
\label{taylor}
\end{equation}

The boundary condition $v^\perp(z=0)=0$, implies $\delta\phi_{(1)}=0$. With the continuity equation, we find :
\vspace{-0.2cm}
\begin{equation}
\bnabla^2_\parallel\delta\phi_{(0)}+\delta\phi_{(2)}+...=0
\label{contrainte}
\end{equation}

We assume that the depth $h$ is much longer than the wavelength $\lambda$ of the free surface perturbation. Hence, the higher-order terms in the Taylor expansion are suppressed by powers of $h/\lambda\ll1$ since we have $\nabla_\|^2={\cal O}(1/\lambda^2)$.
Keeping only the two lowest terms in the Taylor series, we get:
\vspace{-0.2cm}
\begin{equation}
 \begin{array}{rl}
   \delta v^\perp(z) & = \bnabla_\perp \delta \phi \\
                     & =\bnabla_\perp\left(\delta \phi_{(0)}+\frac{z^2}{2}\delta\phi_{(2)}\right)\\
                     & =z\delta\phi_{(2)}\\
    \end{array}
\end{equation}

At the free surface $z=h$ and using (\ref{contrainte}), we then find:
\vspace{-0.2cm}
\begin{equation}
\delta v^\perp(z=h)=-h\bnabla^2_\parallel \delta \phi_{(0)}
\label{deltav}
\end{equation}

In order to find a wave equation for $\delta \phi_{(0)}$, let us take the partial derivative with respect to time $t$ of the equation (\ref{bernoullipert}), then, with the help of the boundary conditions $\delta p(z=h)$, $\delta v^\perp(z=h)$ and of the equation (\ref{deltav}), we subsitute $\delta \dot{h}$ with an equivalent expression. Finally, we end  up with:
\vspace{-0.2cm}
\begin{equation}
\partial_t^2(\delta\phi)+2(\v^\parallel_B.\bnabla_\parallel)\partial_t(\delta \phi)+(\v_B^\parallel\otimes\v_B^\parallel-gh)\bnabla^2(\delta\phi)=0
\label{ondedeltaphi}
\end{equation}
that is the so-called Beltrami-Laplace equation:
\vspace{-0.2cm}
\begin{equation}
\Box \delta \phi_{(0)}=\frac{1}{\sqrt{-g}}\partial_\mu(\sqrt{-g}g^{\mu\nu}\partial_\nu \delta \phi_{(0)})=0
\end{equation}
 provided that we identify $g^{\mu\nu}$ as an inverse metric expressed in the following matrix form:
\vspace{-0.2cm}
\begin{equation}
g^{\mu\nu}=
\begin{pmatrix}
1&\vdots&\v^\parallel_B\\
\ldots\ldots\ldots&.&\ldots\ldots\ldots\ldots\\
\v^\parallel_B&\vdots&\v^{\parallel 2}_B-ghI
\end{pmatrix}
\end{equation}

Using $g^{\mu\nu}g_{\mu\sigma}=\delta^\nu_\sigma$ we get the so-called acoustic metric $g_{\mu\nu}$ in its typical Painlev\'e-Gullstrand form:
\vspace{-0.2cm}
\begin{equation}
g_{\mu\nu}=\frac{1}{c^2}
\begin{pmatrix}
gh-\v^{\parallel 2}_B&\vdots&\v^\parallel_B\\
\ldots\ldots\ldots&.&\ldots\ldots\ldots\ldots\\
\v^\parallel_B&\vdots&-1
\end{pmatrix}
\end{equation}
with $c=\sqrt{gh}$ the velocity of water waves in shallow water which is the analogue of the velocity of light.

It is straightforward to show that the dispersion relation associated to equation  (\ref{ondedeltaphi}) is:
\vspace{-0.2cm}
\begin{equation}
\left(\omega-{\bf k.U}\right)^2\approx c^2k^2
\end{equation}
which describes the propagation of long water waves on a given flow $\U=\v_B$ where we insist on its approximate nature ($kh<<1$). The flow induces a Doppler shift of the angular frequency: we refer the reader to the hydrodynamics literature where the effect of a current on water waves has been discussed extensively \cite{Peregrine, Hedges, Jonsson, Thomas, FS, Lavrenov}. One speaks of blue-shifting (red-shifting) when the current encounters (follows) the waves and the wavenumber increases (decreases). When the flow vanishes, the Beltrami-Laplace operator reduces to the usual d'Alembertian operator.

\begin{figure}[!htbp]
\includegraphics[width=14cm]{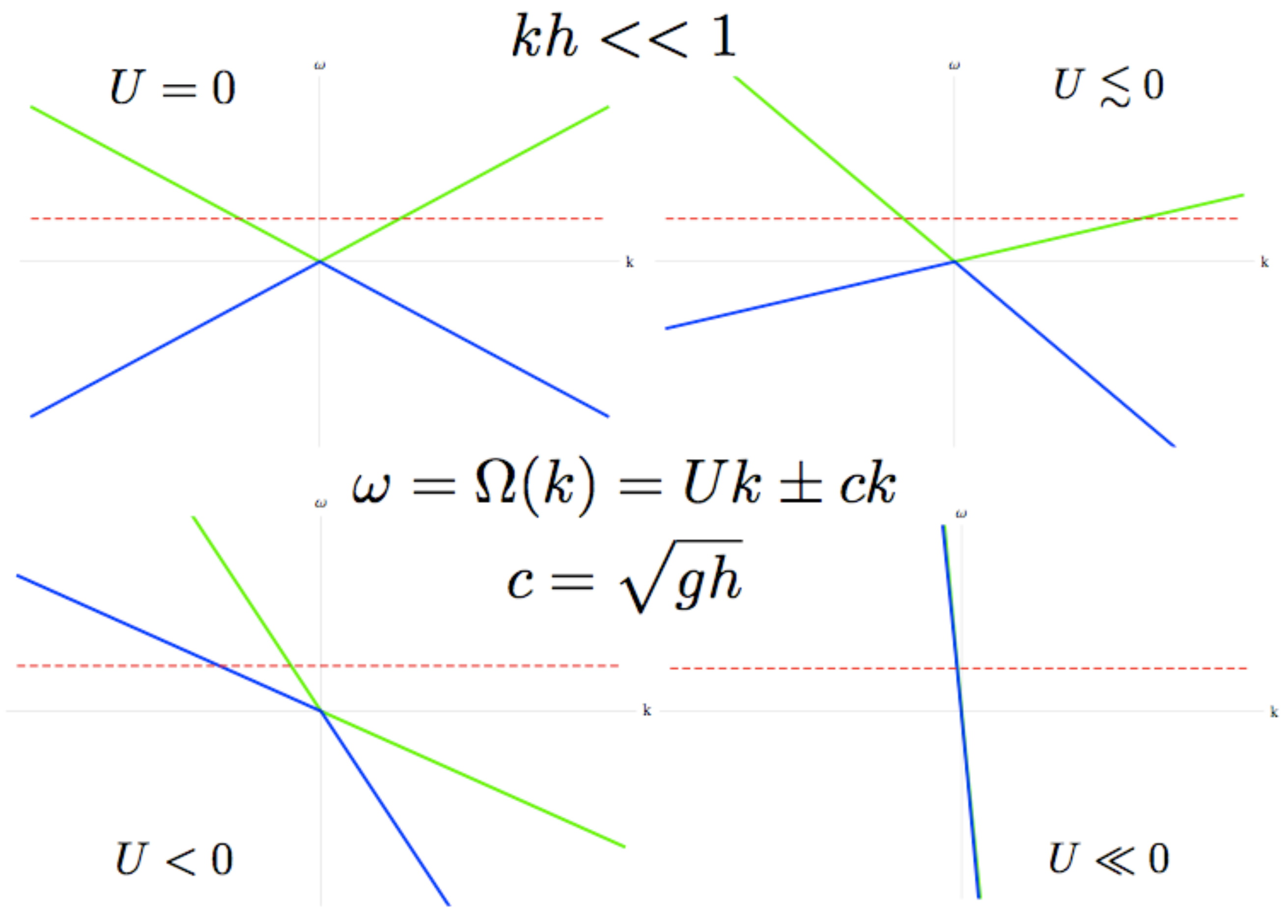}
\caption{Graphical solutions of the dispersion relation $\left(\omega-{\bf k.U}\right)^2\approx c^2k^2$: $\omega =\Omega (k)$ is plotted as a function of $k$ for increasing modulus of the background flow $U<0$. The conserved frequency $\omega$ is the horizontal red dotted line. The green (blue) color corresponds to the positive (negative) branches.}
\label{dispsonic}
\end{figure}

The dispersion relation is solved graphically in Figure \ref{dispsonic}. The so-called transplanckian problem arises when $U=-c$ that is when the wavenumber of the positive solution (in green) diverges at $+\infty$, disappears and then reappears as a new diverging negative solution (in blue) at $-\infty$ for increasing modulus of the flow velocity.

\newpage

\section{Fluid particles' trajectories}

In this part, we recall (without demonstration) some classical results from water waves theory on the trajectories of the fluid particles beneath a water wave \cite{Mei}. For small wave amplitudes ($ka<<1$), the non-linear terms of the Euler equation and of the boundary conditions can be neglected. In 1845, G.B. Airy derived within this approximation the fluid particles' trajectories compatible with the following dispersion relation valid for pure gravity waves without a background flow for a given depth:

\begin{equation}
\omega ^2=gk\tanh{(kh)}
\end{equation}

We denote $z'=0$ the mean water depth of the free surface without wave. Let us consider the following perturbation with respect to rest:
\begin{equation}
z'=\eta (x,t)=a\mathrm{sin}(\omega t-kx)
\end{equation}

Airy computed the resulting velocity profile:
\begin{equation}
u(x,z',t)=a\omega \frac{\mathrm{cosh}(k(z'+h))}{\mathrm{sinh}(kh)}\mathrm{sin}(\omega t-kx)
\end{equation}
and
\begin{equation}
w(x,z',t)=a\omega \frac{\mathrm{sinh}(k(z'+h))}{\mathrm{sinh}(kh)}\mathrm{cos}(\omega t-kx)
\end{equation}
where $u$ and $w$ correspond to the projections of the perturbation velocity in the horizontal and vertical directions.

Under the hypothesis of small displacements, one deduces the horizontal motion of fluid particles:
\begin{equation}
X(x,z',t)-X_{0}=-a\frac{\mathrm{cosh}(k(z'+h))}{\mathrm{sinh}(kh)}\mathrm{cos}(\omega t-kx)
\end{equation}
as well as the vertical motion:
\begin{equation}
Z(x,z',t)-Z_{0}=a\frac{\mathrm{sinh}(k(z'+h))}{\mathrm{sinh}(kh)}\mathrm{sin}(\omega t-kx)
\end{equation}

\begin{figure}[!htbp]
\begin{center}
\includegraphics[width=8cm]{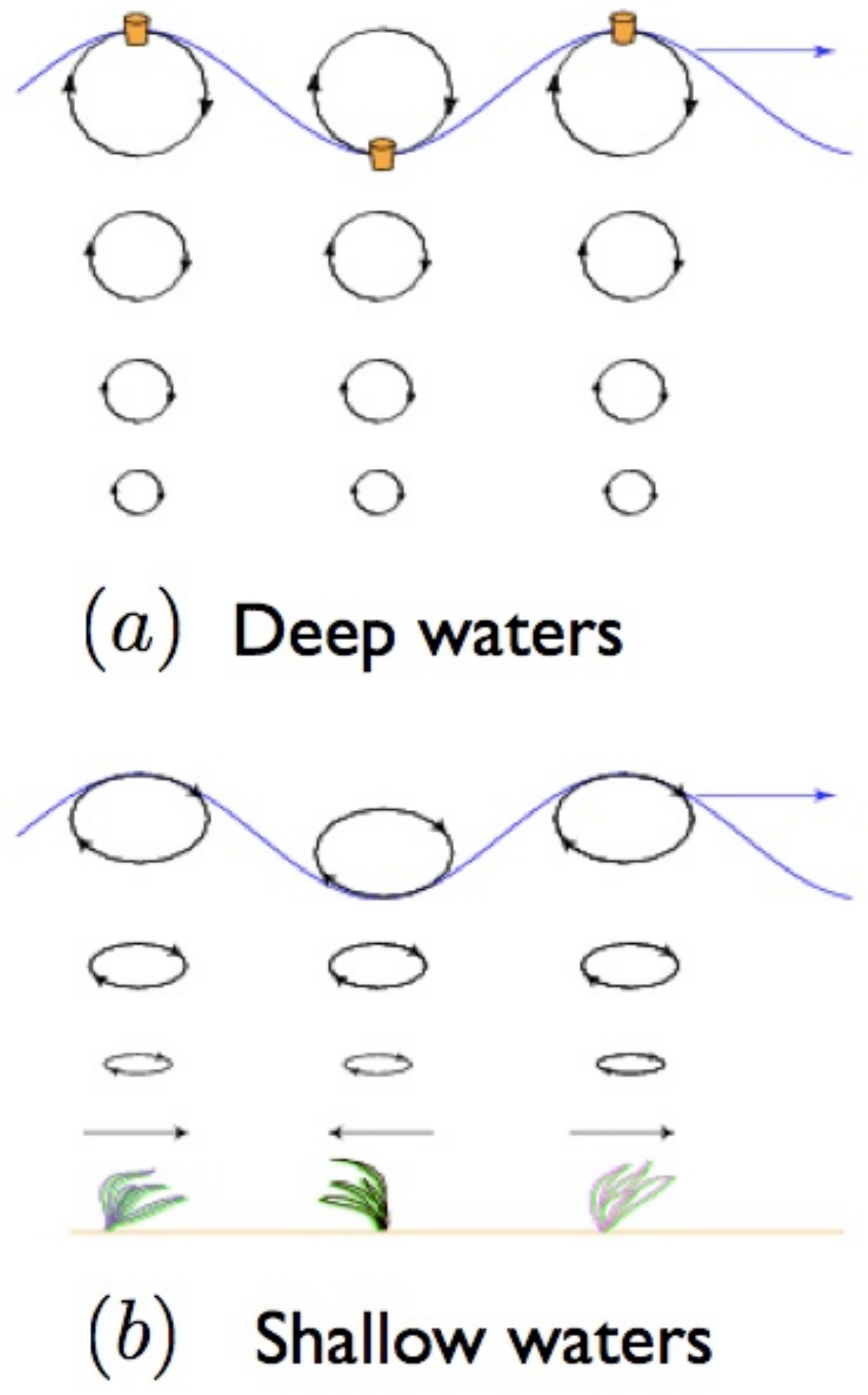}
\caption{The flow generated beneath a surface waves: (a) deep water case; (b) shallow water case.} \label{tray}
\end{center}
\end{figure}

In deep waters (far from the sea shore for example), the fluid particles' trajectories are circular with radius $R$. In shallow waters (close to the beach!), the trajectories flatten and the particles follow an ellipse of semi-axis A and B for respectively the horizontal and vertical motions. In practice, practitioners distinguish three zones :
\begin{itemize}
\item the deep water case ($h/\lambda >1/2$) :

\begin{equation}
A\sim B\sim ae^{kz'}
\end{equation}

The trajectories are circles of radius $R\sim A\sim B$ which decreases exponentially with the depth $z'$.

\item the intermediate case ($1/20<h/\lambda<1/2$) :

\begin{equation}
A=a\frac{\mathrm{cosh}(k(z'+h))}{\mathrm{sinh}(kh)}
\end{equation}
and
\begin{equation}
B=a\frac{\mathrm{sinh}(k(z'+h))}{\mathrm{sinh}(kh)}
\end{equation}

The trajectories are ellipses whose semi-axes diminish with depth. The decrease is slower than the exponential one of deep waters. 

\item the shallow water case ($h/\lambda <1/20$) :

\begin{equation}
A\sim \frac{a}{kh}
\end{equation}
and
\begin{equation}
B\sim \frac{ak(z'+h)}{kh}
\end{equation}

The trajectories are ellipses whose major semi-axis $A$ is independent of the water depth $z'$ and whose minor semi-axis $B$ decreases linearly with $z'$. On the bottom ($z'=-h$), $B$ cancels and the trajectories become a horizontal oscillation of amplitude $A$.

\end{itemize}

In the presence of a current ${\bf U}$, the previous expressions for the velocity field keep the same form provided the dependence of the amplitude (and not of the phase) with the angular frequency $\omega$  is replaced by the relative angular frequency $\omega '= \omega - {\bf k.U}$ \cite{Peregrine, Hedges, Jonsson, Thomas, FS, Lavrenov, Umeyama}. Of course, $u$ becomes $u'+U$ whereas $w'$ is invariant. The particles' trajectories are thus similar to cycloids whose amplitude decreases with the water depth for waves following the current \cite{Umeyama}.

\section{A plethora of dispersive effects}

One of the salient effects of analogue gravity is the possibility to solve the transplanckian problem thanks to the introduction of dispersion close to the horizon of an artificial black hole. As a matter of fact, a major drawback of the original calculation by Stephen Hawking of the black hole radiation is the necessity for the field to have a wavelength which goes to zero as one gets close to the event horizon. Water waves provide several regularization scales in a cascade such as the water depth, the capillary length or even a viscous scale in order to cope with a diverging wavenumber by counter-acting the continuous blue-shifting of the flow...

Let us consider the propagation of gravity waves (without surface tension $\gamma = 0$ for the moment) on a linear shear flow $U(z)=U_0+\Omega z$ with constant plug flow $U_0$ and constant vorticity $\Omega$. Here, one assumes that both the bottom depth and the flow velocity vary slowly such that $\frac{h}{\frac{dh}{dx}}>>\lambda$ and $\frac{U}{\frac{dU}{dx}}>>\lambda$. The dispersion relation between the frequency $\frac{\omega}{2\pi}$ and the wavenumber $k$ writes either with its implicit expression due to Thompson \cite{Thompson}:
\begin{equation}
\left(\omega-U_0k\right)^2 = [gk-\Omega (\omega -kU_0)]\tanh{(kh)}
\end{equation}
or with its explicit expression due to Biesel \cite{Biesel}:
\begin{equation}
\omega = U_0 k -\frac{\Omega}{2}\tanh{(kh)}\pm \sqrt{\left( \frac{\Omega}{2}\tanh{(kh)} \right)^2 +gk\tanh{(kh)}}
\end{equation}

With surface tension, Huang has derived recently the following dispersion relation with its implicit expression  \cite{Huang}:
\begin{equation}
(\omega -kU_0)^2=[gk+\frac{\gamma}{\rho}k^3-\Omega (\omega-kU_0)]\tanh{(kh)}
\end{equation}
which can be written explicitly according to Choi \cite{Choi} in the form:
\begin{equation}
\omega = U_0 k -\frac{\Omega}{2}\tanh{(kh)}\pm \sqrt{\left( \frac{\Omega}{2}\tanh{(kh)} \right)^2 +\left(gk+\frac{\gamma}{\rho}k^3\right)\tanh{(kh)}}
\end{equation}

It is interesting to notice that the "relativistic" dispersion relation $(\omega-U.k)^2=c^2k^2$ is recovered in the long wavelength limit $kh\ll1$ whatever is the dispersive correction. It obvious when dealing with the surface tension since the capillary length is smaller that the long wavelength. It is less obvious for the dispersive effect of vorticity. Indeed, the long wavelength approximation of the Biesel's dispersion relation writes:
\begin{equation}
\omega \simeq U_0 k -\frac{h\Omega}{2}k \pm \sqrt{ghk^2+\frac{\Omega^2 h^2}{4}k^2}
\end{equation}

Fortunately, it can be transformed into the usual dispersion relation associated to the acoustic metric $(\omega-U'.k)^2=c'^2k^2$ provided one introduces renormalized flow and waves velocities $U'=U_0-\Omega h/2$ and $c'=\sqrt{gh+\Omega ^2 h^2/4}$.

Assuming a uniform flow in the vertical direction ($\Omega =0$), the dispersion relation becomes \cite{Peregrine, Hedges, Jonsson, Thomas, FS, Lavrenov}:
\begin{equation}
(\omega - Uk)^2 \simeq \left(gk+\frac{\gamma}{\rho}k^3\right)\tanh (kh)
\end{equation}

\begin{itemize}
\item
In the shallow water limit $kh<<1$,
\begin{equation}
(\omega-Uk)^2 \simeq ghk^2+\left(\frac{\gamma h}{\rho}-\frac{g h^3}
{3}\right)k^4+\mathcal{O}(k^6)
\end{equation}
the dispersion relation is identical to a BEC-type phonons spectrum:
\begin{equation}
(\omega-Uk)^2 \simeq c^2k^2\pm c^2\xi^2k^4
\end{equation}
with the corresponding "healing length":
\begin{equation}
\xi = \sqrt{\left\vert l_c^2-\frac{h^2}{3}\right\vert }
\end{equation}
An  interesting observation is that the superluminal correction can have a negative sign in contrast to the BEC case if the capillary length is less than $h/\sqrt{3}$ or even null...

\item
In the deep water limit $kh>>1$, the dispersion relation looses its "relativistic/acoustic" branch:
\begin{equation}
(\omega - Uk)^2 \simeq gk+\frac{\gamma}{\rho}k^3
\end{equation}
\end{itemize}

Viscosity has both a dissipative (imaginary term) and a dispersive (real term) contributions to the dispersion relation. By dimensional analysis, it is obvious that the typical viscous scale would be of the order of $\delta \approx \sqrt{\frac{\nu}{\omega}}$ otherwise known as the Stokes viscous length which is the scale of the viscous boundary layer \cite{Mei}.

\section{Hydrodynamic Horizons}

In this part, we propose a generalized definition of a horizon with respect to the usual custom in General Relativity.  Condensed matter horizons and here, hydrodynamic horizons lead to a dispersive-like definition. What is a Horizon ? The word horizon derives from the Greek ``$o\rho \iota \zeta \omega \nu$ $\kappa \upsilon \kappa \lambda o \varsigma$" ($horizon$ $kyklos$), ``separating circle", from the verb ``$o\rho \iota \zeta \omega$" ($horizo$), ``to divide, to separate", from the word ``$o\rho o \varsigma$" (oros), ``boundary, landmark". In the Fable of Jean de la Fontaine recalled at the beginning of this chapter, will the Lamb be right to argue against the Wolf that the waves he creates as he is drinking at the river border will not climb against the current and reach the Wolf ? Will a frontier separate the Lamb from the Wolf: will a horizon form ? Will the position of the frontier depend on the period of the waves: will the horizon be dispersive or not ?

\subsection{Non-Dispersive Horizons}

The analogy between the propagation of light in a curved space-time and the propagation of long gravity waves on a current features the so-called "acoustic/relativistic" dispersion relation $\left(\omega-{\bf k.U}\right)^2\approx c^2k^2$ as a common characteristic for both systems assuming $kh <<1$ and without surface tension. A simple dimensional analysis of it:
\begin{equation}
\omega ^2 \approx U^2k^2 \approx ghk^2
\end{equation}
allows to infer scaling laws for the wavenumber:
\begin{equation}
k \approx \frac{\omega}{\sqrt{gh}}
\end{equation}
and the blocking velocity:
\begin{equation}
U \approx \sqrt{gh}
\end{equation}
which we confirm by solving the ``relativistic" dispersion relation as a polynomial in $k$:
\begin{equation}
k_h= \frac{\omega}{U+\sqrt{gh}}
\end{equation}
implying the transplanckian problem ($k_h \to \infty$) when the blocking velocity ($U^*$ in modulus) of long gravity waves matches the current flow:
\begin{equation}
U_h= -\sqrt{gh}
\end{equation}

\begin{figure}[!htbp]
\includegraphics[width=14cm]{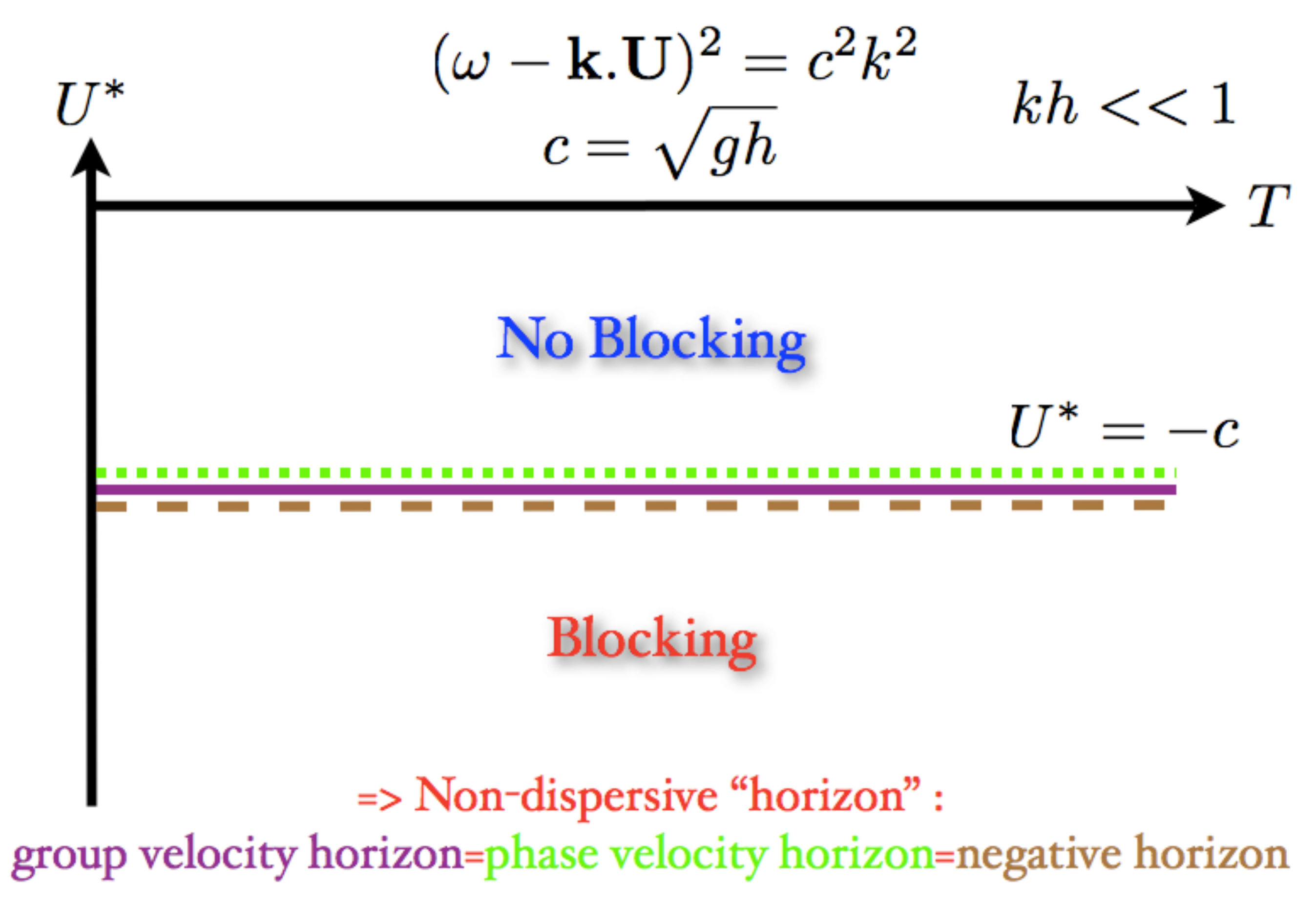}
\caption{Phase-Space -U versus T- of the dispersion relation $\left(\omega-{\bf k.U}\right)^2\approx c^2k^2$.}
\label{3H}
\end{figure}

This non-dispersive definition of a horizon in hydrodynamics corresponds to the definition of General Relativity where the pure temporal matrix element of the Painlev\'e-Gullstrand metric is cancelled:
\begin{equation}
g_{00}=0
\end{equation}
leading to:
\begin{equation}
U=-c=-\sqrt{gh}
\end{equation}
Here, it is crucial to understand that the relativistic horizon hides in fact three intricate horizons (Figure \ref{3H}: blocking velocity $U^*$ versus the wave period $T$): a group velocity horizon ($c_g=\frac{\partial \omega}{\partial k}=U+c=0$), a phase velocity horizon ($c_\phi =\frac{\omega}{k}=U+c=0$) and a negative horizon (or negative energy mode horizon): negative relative frequencies $\omega-{\bf k.U}<0$ can appear. This last fact implies that Stimulated Hawking Radiation can be observed in Classical Physics using water waves and this is one of the major interests of the analogue gravity program for the Fluid Mechanics community \cite{NJP08, PRL09, NJP10, Silke, PRE11}. As soon as there is a phase velocity horizon, this one is identical with a negative horizon. Because of dispersion, a phase velocity horizon can be absent whereas a negative group velocity horizon can be present (see below).

These ``negative energy waves" are well known in Hydrodynamics. Werner Heisenberg discovered them in his PhD Thesis on the stability of the plane Couette flow. He showed that viscosity can have a destabilizing effect if negative energy waves (also named Tollmien-Schlichting waves) are present (at the so-called critical layer corresponding to a phase velocity horizon) in a unidirectional non-inflectional plane flow which is normally stable if inviscid according to the classical Rayleigh criterion \cite{Darrigol, FS}!

\subsection{Dispersive Horizons}

\begin{figure}[!htbp]
\includegraphics[width=14cm]{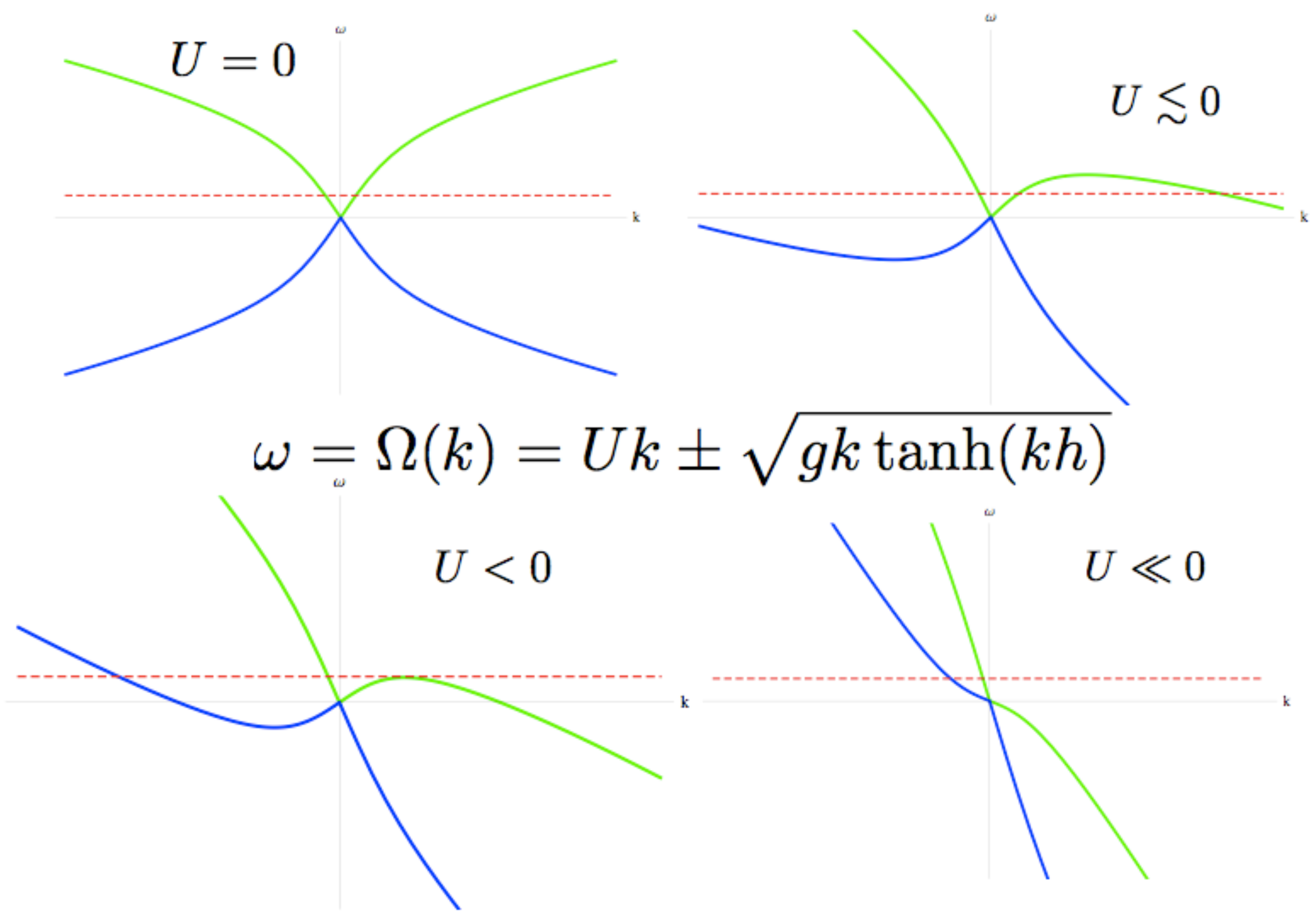}
\caption{Graphical solutions of the dispersion relation $\left(\omega-{\bf k.U}\right)^2=gk\tanh(kh)$. The conserved frequency $\omega$ is the horizontal red dotted line. The green (blue) color corresponds to the positive (negative) branches.}
\label{dispth}
\end{figure}

How is the definition of a horizon modified in the presence of dispersion ? Wave blocking is a process where a flow separates a free surface into a flat and a deformed surface. The boundary defines a ``horizon". A wave phenomenon implies the existence of a  dispersion relation $\omega=\Omega (k)$. At the boundary, the energy flow of the system ``waves+current" cancels:
\begin{equation}
c_{group}^{wave+current}=\frac{\partial \Omega}{\partial k}=0
\end{equation}
This last criterion will define a hydrodynamic horizon as a group velocity horizon (or turning point using WKBJ terminology). Of course, one recovers the non-dispersive definition $U=-c$ for an ``acoustic/relativistic" dispersion relation. We treat here the simple case of a white hole horizon which is the time reverse of a black hole horizon. As previously, the dispersion relation for water waves in arbitrary depth is solved graphically (Figure \ref{dispth}). An extremum of the function $\omega = \Omega (k)=Uk\pm \sqrt{gk\tanh(kh)}$ corresponds to a horizon.

Some scaling laws can be derived in the high dispersive regime where $kh >>1$:
\begin{itemize}
\item without surface tension. Dimensional analysis leads to:
\begin{equation}
\omega ^2 \approx U^2k^2 \approx gk
\end{equation}
that is :
\begin{equation}
k \approx \frac{\omega ^2}{g}
\end{equation}
and:
\begin{equation}
U \approx \frac{\omega}{k}\approx \frac{g}{\omega} \approx gT
\end{equation}
The rigorous mathematical treatment gives \cite{PRL09}:
\begin{equation}
k_g=\frac{4\omega^2}{g}
\end{equation}
and
\begin{equation}
U_g=-\frac{g}{4\omega}=-\frac{gT}{8\pi}
\end{equation}
The blocking velocity $U^*$ depends now on the incoming period of the water waves (Figure \ref{2range}). Depending on the period, we have either $U^*=U_h$ for long waves or $U^*=U_g$ for short waves.

\begin{figure}[!htbp]
\includegraphics[width=14cm]{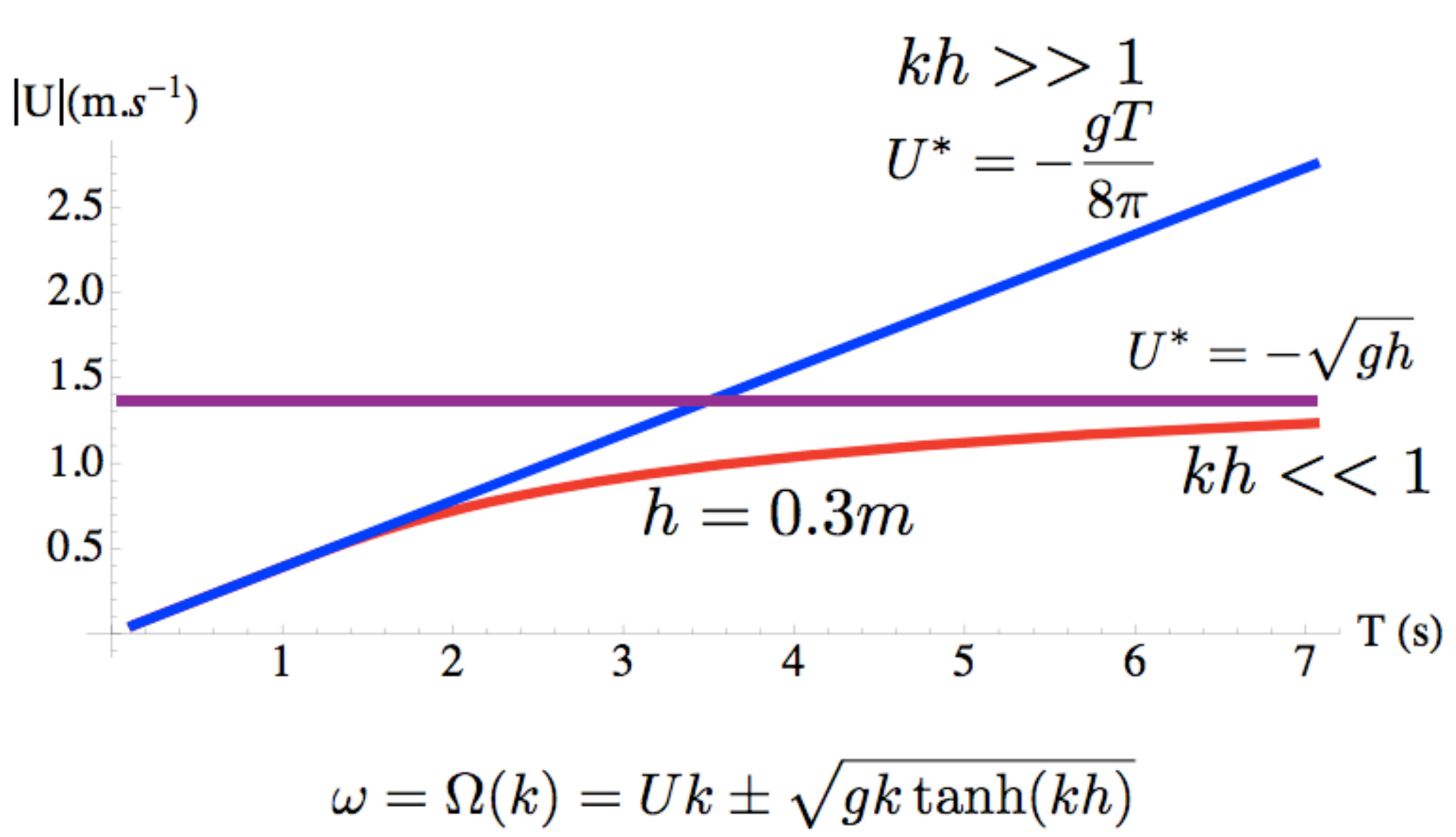}
\caption{Superposed Phase-Spaces -U versus T- for the shallow and deep water cases without surface tension.}
\label{2range}
\end{figure}

\item with surface tension. Dimensional analysis leads to:
\begin{equation}
\omega ^2 \approx U^2k^2 \approx gk \approx \frac{\gamma}{\rho}k^3
\end{equation}
that is:
\begin{equation}
k\approx \left(\frac{\rho g}{\gamma}\right)^{1/2}
\end{equation}
and:
\begin{equation}
U \approx \left(\frac{\gamma g}{\rho}\right)^{1/4}
\end{equation}
The rigorous mathematical treatment gives \cite{NJP10}:
\begin{equation}
k_\gamma= \left(\frac{\rho g}{\gamma}\right)^{1/2}
\end{equation}
and:
\begin{equation}
U_{\gamma}=-\sqrt{2}\left(\frac{\gamma g}{\rho} \right)^{1/4}
\end{equation}

A new horizon (in fact two) appears. Blue-shifted waves and negative energy waves can be reflected at a blue horizon and a negative horizon whose common asymptotic value is $U_\gamma$ (see \cite{NJP10} for the details and the corresponding chapter in this book). Two maxima and a minimum appear in the graphical analysis of the dispersion relation (Figure \ref{Badulin}). A cusp where the white and blue horizons merge appears in the Phase-Space (Figure \ref{PS}).

\begin{figure}[!htbp]
\includegraphics[width=14cm]{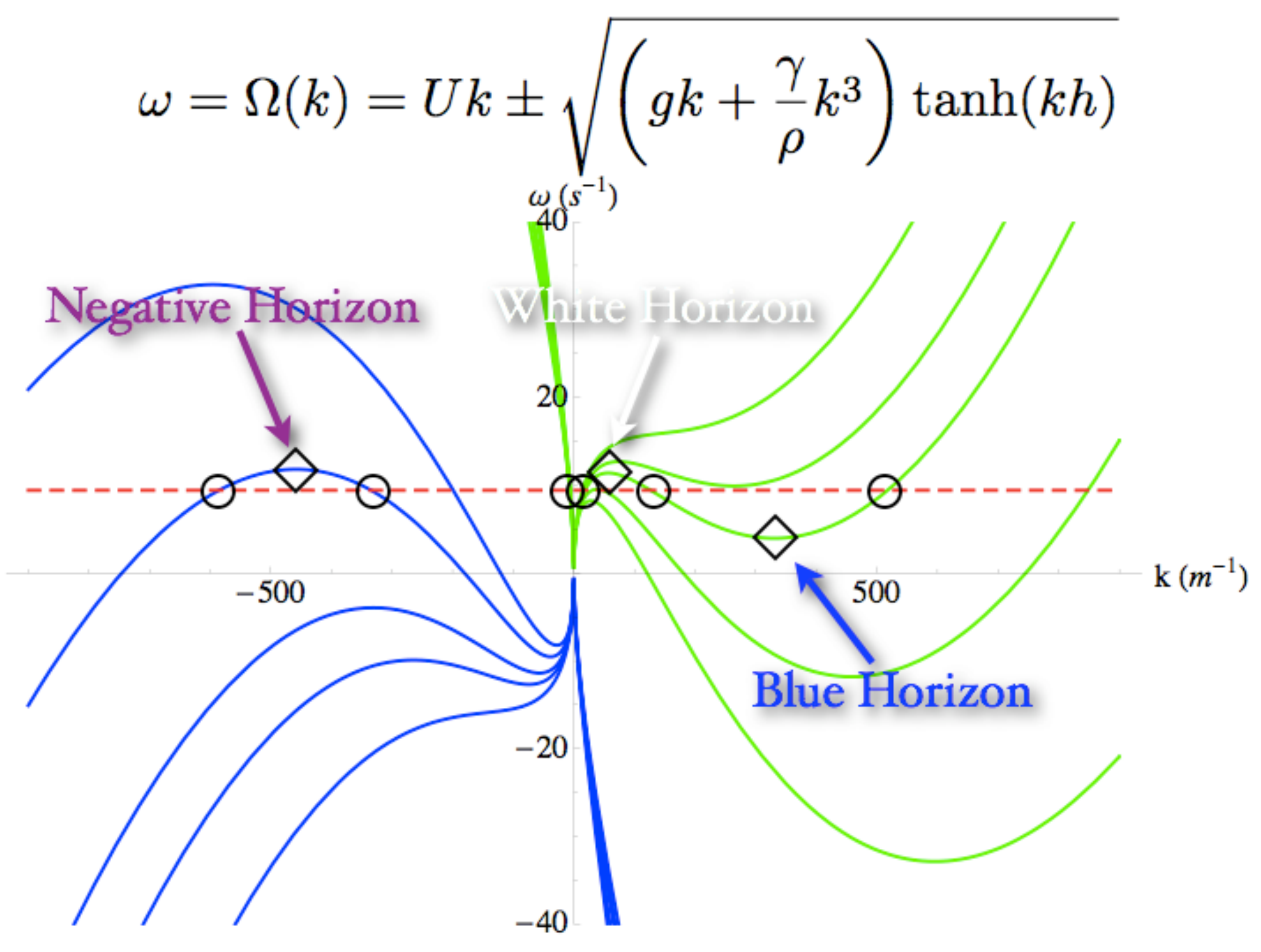}
\caption{Graphical solutions of the dispersion relation $\left(\omega-{\bf k.U}\right)^2= \left(gk+\frac{\gamma}{\rho}k^3\right)\tanh(kh)$. The conserved frequency $\omega$ is the horizontal red dotted line. The green (blue) color corresponds to the positive (negative) branches.}
\label{Badulin}
\end{figure}

\begin{figure}[!htbp]
\includegraphics[width=14cm]{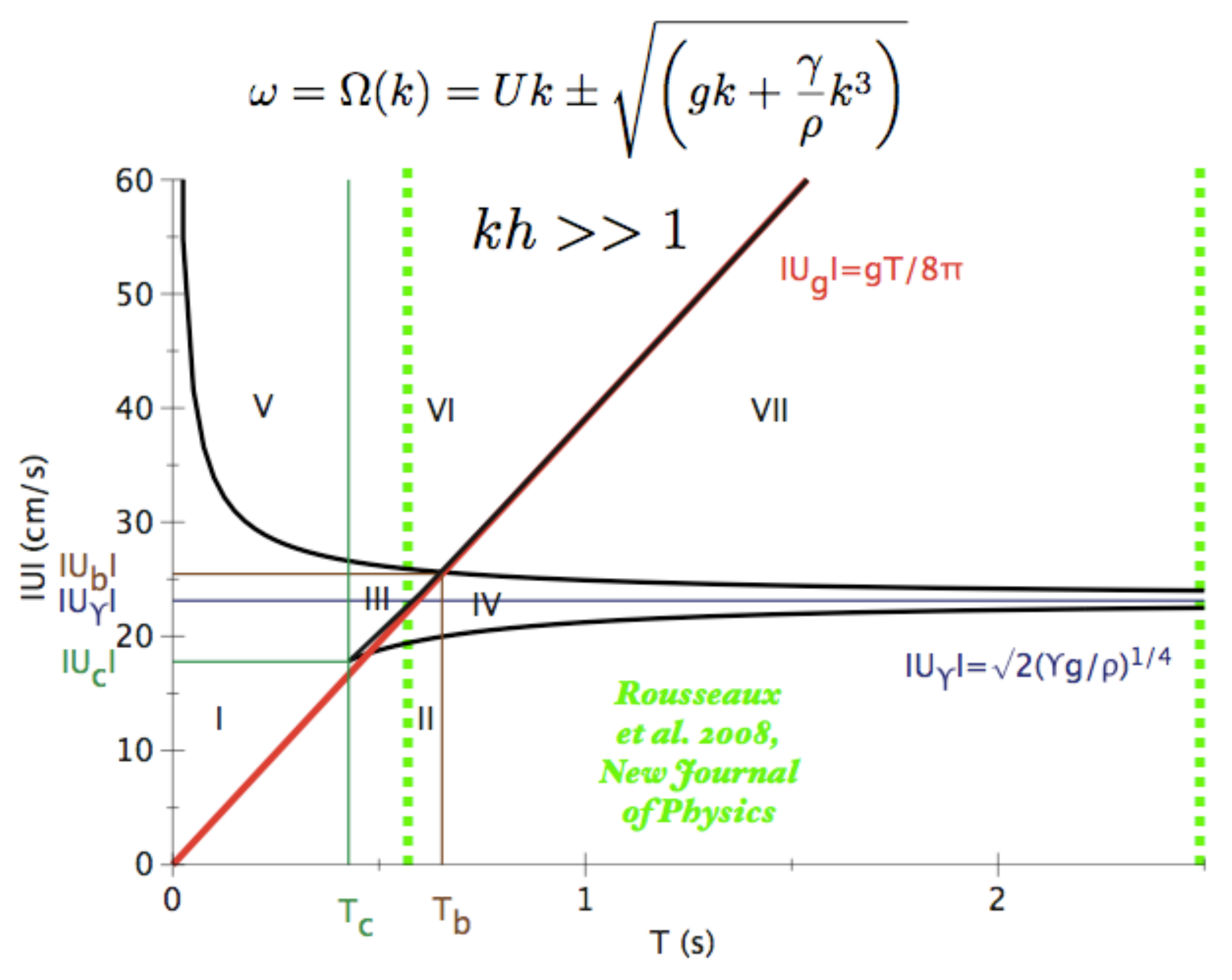}
\caption{Phase-Space -U versus T- for the deep water case including surface tension.}
\label{PS}
\end{figure}

When the water depth changes, the dispersion relation is either $(\omega-Uk)^2 \simeq c^2k^2\pm c^2\xi^2k^4$ for $kh<<1$ allowing dispersive corrections (only a negative horizon remains with the positive quartic correction) or $\left(\omega-{\bf k.U}\right)^2= \left(gk+\frac{\gamma}{\rho}k^3\right)\tanh(kh)$ and three horizons are observed (Figure \ref{heffect}). $h^*=2l_c$ determines the transition depth between both behaviors.

\begin{figure}[!htbp]
\includegraphics[width=14cm]{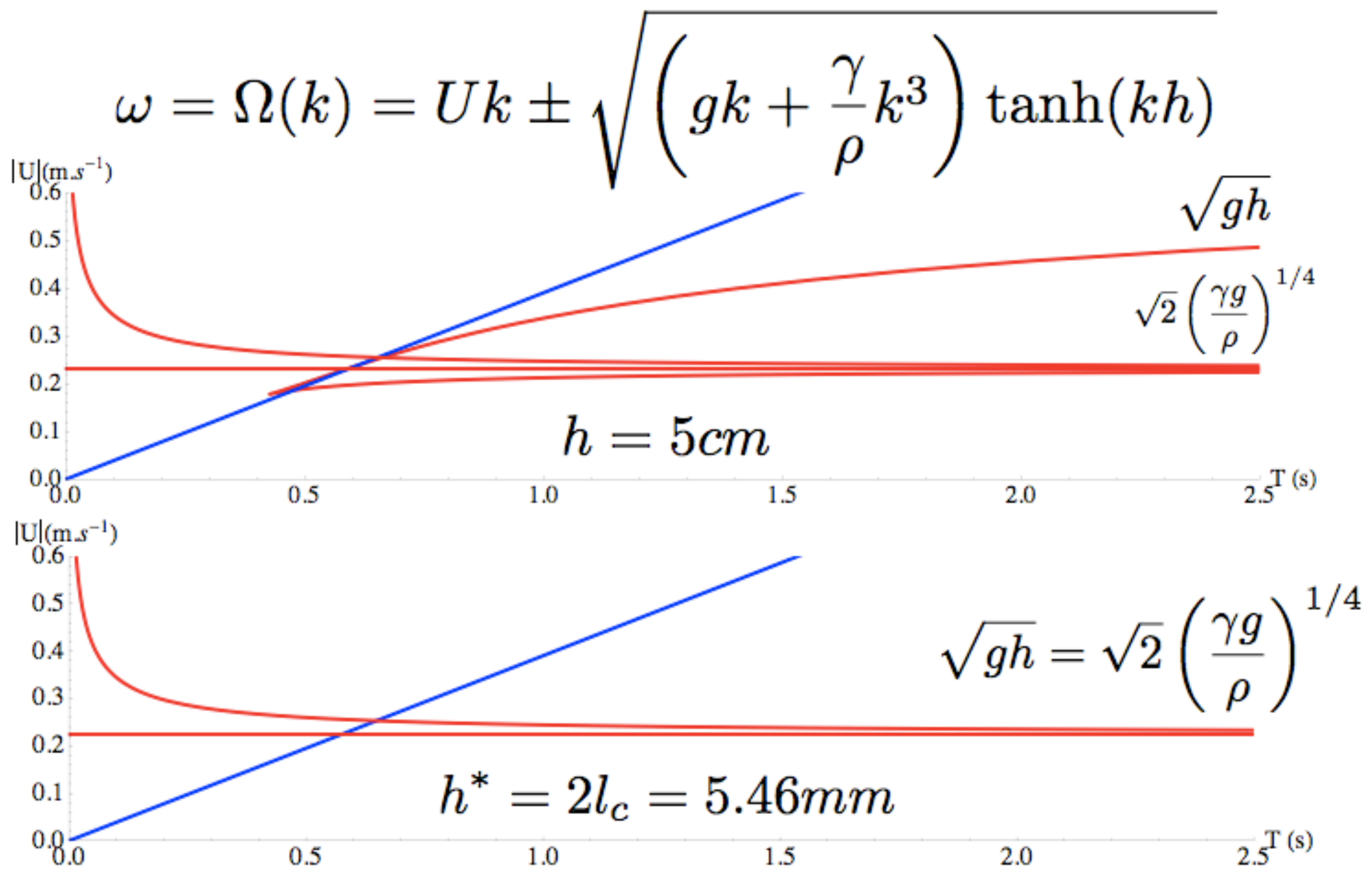}
\caption{Phase-Spaces -U versus T- for a changing water depth including surface tension.}
\label{heffect}
\end{figure}

\newpage

\item with vorticity. Similar arguments would lead to a new horizon replacing $U_\gamma$ when including vorticity $\Omega$ with qualitatively the same behavior in the limit $kh>>1$:
\begin{equation}
k_{\Omega} \approx \left(\frac{\rho \Omega ^2}{\gamma}\right)^{1/3}
\end{equation}
and
\begin{equation}
U_\Omega \approx \left(\frac{\Omega \gamma}{\rho}\right)^{1/3}
\end{equation} 

\end{itemize}

\subsection{Natural and Artificial Horizons}

\begin{figure}[!htbp]
\includegraphics[width=14cm]{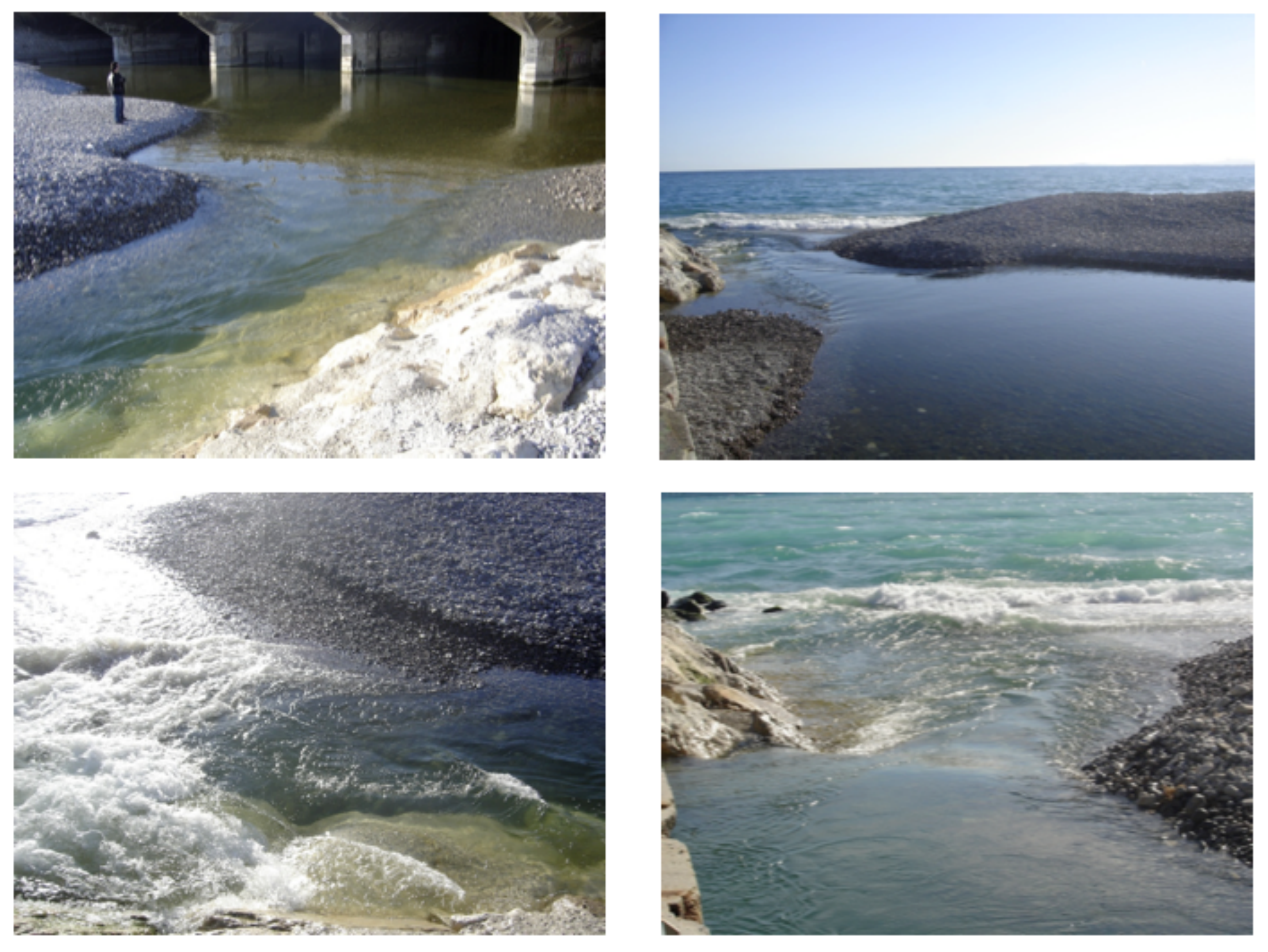}
\caption{A natural white hole of the French Riviera.}
\label{mouth}
\end{figure}

In this part, we give some examples of water wave horizons. In hydrodynamics, white holes are more usual than black holes whose canonical example is the draining flow in the bathtub. A river mouth dying in the sea is a nice case of a natural white hole: the sea waves are blocked by the river flow. Figure \ref{mouth} is an example found by the author when he used to walk on the Promenade des Anglais in Nice (France). This white hole inspired the following studies \cite{NJP08, NJP10, Silke}.

\begin{figure}[!htbp]
\includegraphics[width=14cm]{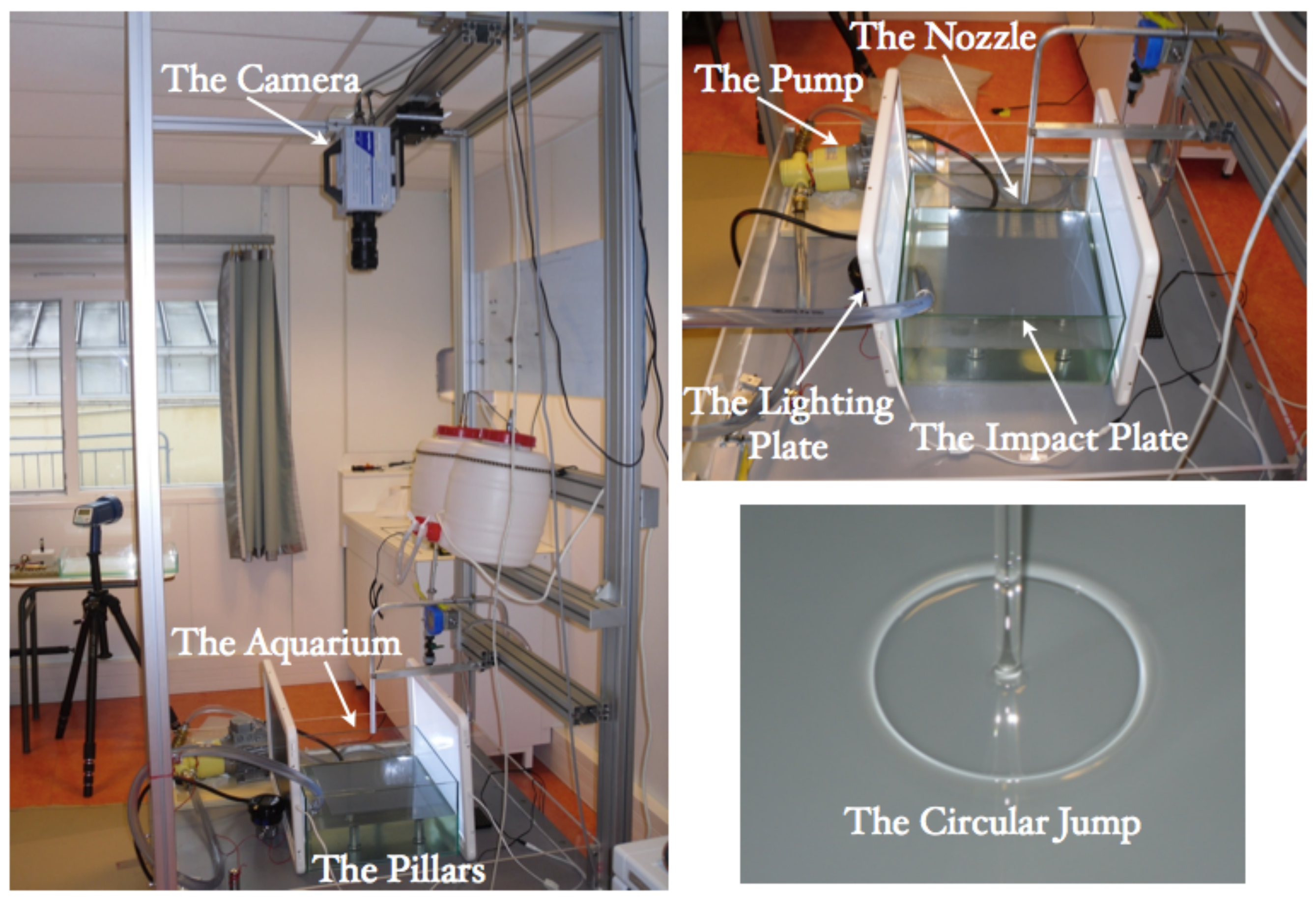}
\caption{A laboratory white hole in the kitchen sink.}
\label{setupJ}
\end{figure}

A more controlled white hole in the laboratory was suggested a few years ago by Volovik \cite{Volovik}: the circular jump in the kitchen sink (Figure \ref{setupJ}). We studied its related Mach cone and its dispersive properties in \cite{PRE11}. 

\begin{figure}[!htbp]
\includegraphics[width=12cm]{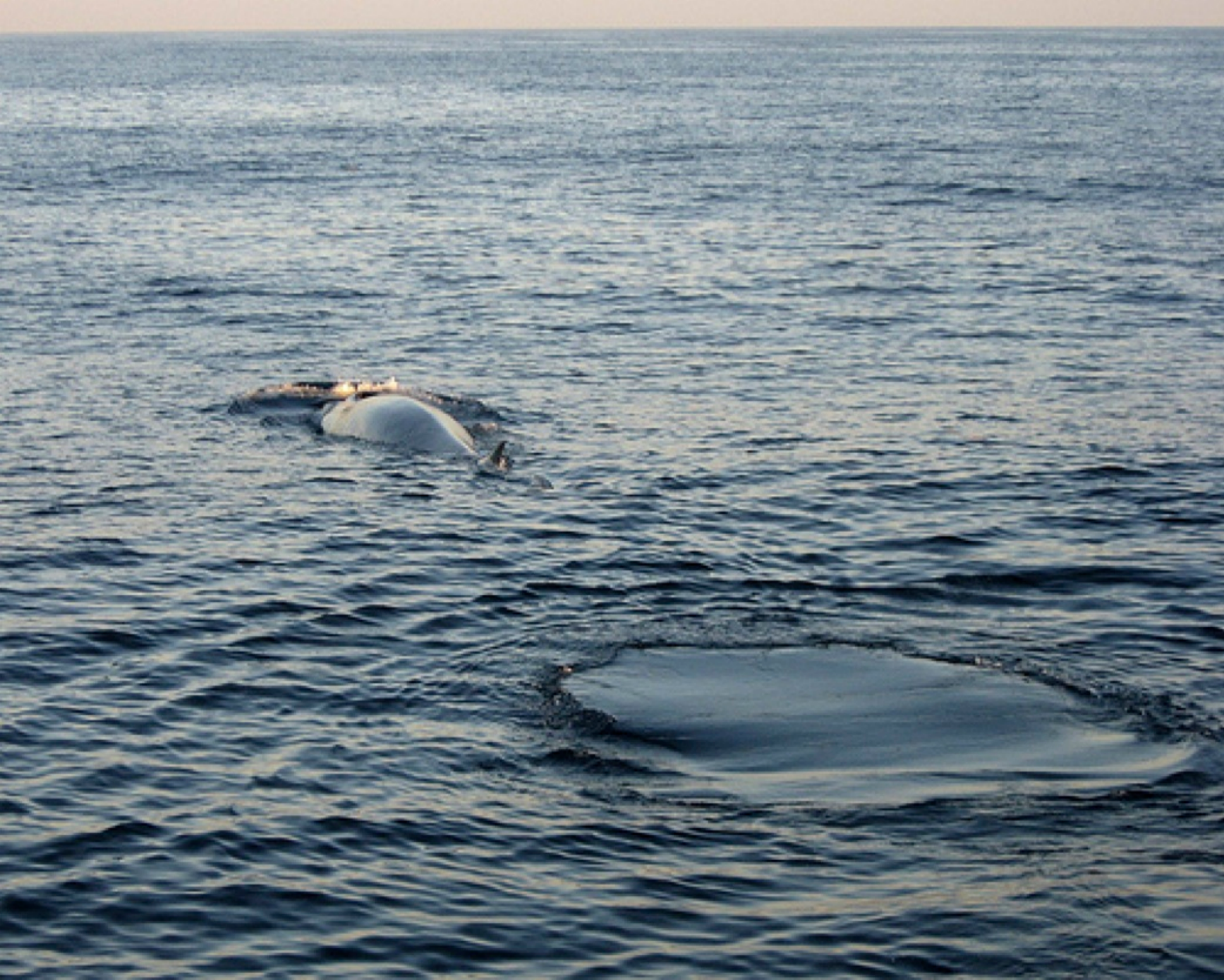}
\caption{A "biological" dispersive white hole.}
\label{Flukeprint}
\end{figure}

Recently, the author became aware of a biological-induced white hole with interesting dispersive properties, namely the whale fluke-print. As a whale swims or dives, it releases a vortex ring behind its fluke at each oscillation. The flow induced on the free surface is directed radially and forms a oval patch that gravity waves cannot enter whereas capillary waves are seen on its boundary (Figure \ref{Flukeprint}). 

Dispersive and non-dispersive horizons are all encountered in nature and can be simulated in the laboratory. Dispersion has another intriguing consequence, namely the appearance of current-induced zero-frequency waves which appear spontaneously. These so-called ``zero modes" have no counterpart in General Relativity so far...

\newpage

\subsection{Zero Modes}

A flat interface can be considered as a wave with zero frequency $\omega = 0$ and zero wavenumber $k=0$.  When a spatially varying current is flowing under such an interface, the infinite wavelength of the interface can be reflected at a blocking line (creation of a group velocity horizon). This process produces a static ($\omega = 0$) jump through the interferences between the incident wave (flat surface with infinite wavelength $k=0$) and the reflected one. This explains the formation of the  circular \cite{Volovik, PRE11} and hydraulic \cite{Unruh08} jumps. An undulation is observed which has a zero phase velocity but a non-zero group velocity and thus withdraws energy from the horizon towards infinity. When surface tension is present, the "gravity" jump is decorated by static capillary ripples inside the circular jump \cite{Rolley}.

\subsubsection{The zero mode (static undulation) for gravity waves}

Two opposite wavenumbers are solutions of the dispersion relation for a zero frequency :
\begin{equation}
(0-Uk)^2= gk\tanh(kh)
\end{equation}

Clearly, there is no threshold since there is always a solution whatever the velocity of the flow. Whatever the water depth, the slightest current flow induces a free surface deformation.

Let us take the extreme shallow waters limit ($kh<<1$):
\begin{equation}
U¼ ^2k^2 \simeq ghk^2
\end{equation}
The threshold velocity for the zero mode appearance would correspond to:
\begin{equation}
U¼\simeq \sqrt{gh}=c_{phase}=c_{group}
\end{equation}
that is:
\begin{equation}
Fr¼=\frac{U¼}{\sqrt{gh}}=1
\end{equation}
in terms of the dimensionless Froude number $Fr$. This latter constraint is well known in Hydraulics as the condition of appearance of the hydraulic jump when water flows over a bump: 
\begin{equation}
Froude=Fr=\frac{U}{\sqrt{gh}}=\frac{U}{c_{phase}}=\frac{U}{c_{group}}=M=Mach
\end{equation}
which is similar to the supersonic-subsonic transition of air flows in aerodynamics described by the so-called Mach number $M$ \cite{PRE11}.

Then, one distinguishes in Hydraulics the following regimes: 
\begin{itemize}
\item
$Fr<1$: (a) subcritical-to-subcritical flow over a bump. A group velocity horizon can appear but no phase velocity horizon \cite{NJP08}. No hydraulic jump is created but a static undulation is observed. The water depth decreases on average over the bump.

\item
$Fr>1$: (b) supercritical-to-supercritical flow over a bump. The water depth increases on average over the bump.
\item
$Fr=1$: (c) subcritical-to-supercritical flow over a bump. The group and phase velocity horizons are the same. A hydraulic jump appears as part of the static undulation \cite{Unruh08, Silke}.

\end{itemize}

\subsubsection{The zero mode (static undulation) for capillo-gravity waves}

If one takes into account the effect of surface tension:
\begin{equation}
(0-U¼k)^2= \left(gk+\frac{\gamma}{\rho}k^3\right)\tanh(kh)
\end{equation}

One is lead to the existence of a velocity threshold which corresponds to the minimum of the phase velocity with the wavenumber $U_\gamma=-\sqrt{2} \left(\frac{\gamma g}{\rho}\right)^{1/4}$ \cite{NJP10}.

For the case of the circular jump assuming $kh<<1$,
\begin{equation}
(0-U¼k)^2= c^2k^2+\left(l_c^2-\frac{h^2}{3}\right)k^4
\end{equation}
the following condition:
\begin{equation}
h<\sqrt{3}l_c \Rightarrow U¼>c=\sqrt{gh}
\end{equation}
implies the existence of static capillary undulations in the supersonic region of the circular jump \cite{Rolley}.

We have seen how the dispersion relation explains the appearance of  a horizon as well as the evolution of the wavelength of the converted modes. How does the amplitude of the modes evolve ?

\section{The "Norm"}

\begin{flushright}
  \begin{minipage}[h]{11cm}
    \begin{quote}
    \footnotesize
   {\it If someone tells you that he knows what $E=\hbar \omega$ means,\\ tell him that he is a liar.}
      \begin{flushright}
Albert Einstein
      \end{flushright}
    \end{quote}
  \end{minipage}
\end{flushright}

In this final part, we will show how the so-called "norm" used by relativists in order to derive the Hawking spectrum is nothing else than the wave action, a pure classical concept.

In 1905, Albert Einstein pointed out that the four-momentum and the four-wave vector transform similarly under a Lorentz boost. This simple remark was fundamental in order to infer the existence of the light quantum whose energy is proportional to the frequency. The factor of proportionality was the Planck constant and Physicists soon realized that the latter constant of nature was measured in units of action. Quantum Mechanics then will shortly take its roots in Analytical Mechanics. In the famous 1911 Solvay conference \cite{Solvay}, Lorentz wondered about the paradoxical behavior of a harmonic oscillator like a pendulum whose frequency was made to change slowly with time by reducing its length. Indeed, the corresponding quantum behavior of the oscillator would forbid a change in the quantum number describing the state of the oscillator since the frequency variation would not be high enough to allow transition to another state. Einstein pointed out that both the energy and the frequency of the pendulum would change with time but not their ratio as discovered by Rayleigh in 1902 \cite{Rayleigh}. Ehrenfest showed that the ratio of energy to frequency namely the action was an ``adiabatic invariant".  Adiabatic invariants of a given dynamical system are approximate constants of motion which are approximately preserved during a process where the parameters of the system change slowly on a time scale, which is supposed to be much larger than any typical dynamical time scale. They are the quantities to quantize when switching from Analytical Mechanics to Quantum Mechanics. A similar relation was discovered later by De Broglie between the momentum and the wavenumber. It should be borne in mind that the ratio of the action to the Planck constant is the number of photons which is another way to interpret the norm (as we will see) as the number of photons/phonons times the quantum of action in a quantum context.

Water waves are an example of a classical field. Thus, we can anticipate that the fluid system will have a corresponding wave action density defined as the ratio between a mean energy density (computed by averaging the instantaneous energy density on a spatial wave period) and the wave frequency. This action $J=E/\omega$ is assumed to be an adiabatic invariant (see \cite{S1} for a demonstration based on classical field theory). Then, if prime denotes a moving frame of reference with velocity {\bf v}, we must have $J'=J$ for a Galilean boost (recall that Einstein dealt with Lorentz transformations applied to light) that is \cite{S2, S3} :
\begin{equation}
\frac{E'}{\omega '}=\frac{E}{\omega}
\end{equation}
which is valid if and only if we have the following transformations :
\begin{equation}
E'=E-{\bf v.P}, \quad {\bf P}'={\bf P}
\end{equation}
and
\begin{equation}
\omega '=\omega-{\bf v.k}, \quad {\bf k}'={\bf k}
\end{equation}
The latter formulae are just the usual Doppler effect whereas the former correspond to the change of energy/momentum for a classical wave and NOT a particle \cite{Carroll}. These would apply to a quasi-particle that is a collective excitation : phonon in acoustics or ripplons for water waves. In the following, momentum and energy would refer to quasi-momentum and quasi-energy if not specified.

It is now obvious that the energy in the moving frame will be given by $E'=E(1-v/c_\phi)$ where $c_\phi =\omega /k$ is the phase velocity in the rest frame \cite{S2,S3}. In order to have negative energy waves ($E'<0$), the so-called Landau criterion must be fulfilled $\omega - {\bf U.k}<0$ since the energy in the rest frame $E>0$ is always positive. It is well known that superfluidity is lost when negative energy waves are created at the minimum of the roton spectrum \cite{Volovik}. It is similar to waves creation (Cerenkov-like effect) by an object in a flowing current $U$ perforating the interface between water and air (capillary waves in the front and gravity waves in the rear) when $\mathrm{min}(c_\phi )<U$ : the phase velocity $c_\phi =\omega /k$ features a minimum under which no waves are created (cf. Thomson and Helmholtz fishing line as described by Darrigol \cite{Darrigol} and the  corresponding chapter in this book). Let us recall that, in the direct space, the mean energy density (or pseudo-energy) for water waves (without any current) is proportional to the square of the amplitude $E=1/2\rho g a^2$ where $a$ is the amplitude of the wave \cite{Mei}.

Here, we must be careful when we want to evaluate the wave energy in the moving frame of the current because the velocity of the flow is changing with space. Hence, every part of the water waves wavelength will be "desynchronized" by the spatial-dependent Doppler effect due to the current. That is why Weinfurtner et al. \cite{Silke} introduced a time shift $t_c=\int \frac{dx}{U(x)}$ which is reminiscent of Carroll kinematics for classical waves $t'=t-vx/c^2$ and $x'=x$ or $t'=t-x/V_0$ where $V_0=c^2/v$ is the dual velocity associated to the wavefront \cite{Carroll}. The Carrollian time shift writes $dt'=dt-dx/V_0$ in differential form where $V_0$ is now a function of space x in the water waves problem. The $t_c$ coordinate has dimension of time and its associated "wavenumber" $f_c$ has units of a frequency. The usual convective derivative operator $\partial _t + U(x)\partial _x$ becomes $\partial _t +\partial _{t_c}$ and in Fourier transform space $f+f_c$. Then, when analysing data in the Carrollian coordinate system, the amplitude of the wave $\eta $ is a function of both the normal time $t$ and the Carrollian time $t_c$ that is in the Fourier transform space $\tilde{\eta }(f, f_c)$. The wave action density in the Fourier space $\hat{J}_{wave}$ is by definition the integral on the different Carrollian times of the ratio between the Fourier transform of the wave energy density (E) and the Fourier transform of the relative angular frequency ($\omega '$):
\begin{equation}
\hat{J}_{wave}=\int \frac{|\tilde{\eta} (f, f_c)|^2}{f+f_c}df_c
\end{equation}

The expression of $J_{wave}$ is similar to the Zeldovich formula for the number of photons $N$ when dealing with plane electromagnetic waves that are not monochromatic ($\omega = \pm c_L |{\bf k}|$):
\begin{equation}
N=\frac{1}{\hbar}\frac{1}{8\pi}\int d^3k\frac{|\hat{{\bf E}}({\bf k},t)|^2+|\hat{{\bf B}}({\bf k},t)|^2}{\pm c_L|{\bf k}|}
\end{equation}
where $c_L$ is the light velocity and $\hat{{\bf E}}({\bf k},t)$, $\hat{{\bf B}}({\bf k},t)$ are the Fourier transforms of the electric and magnetic fields \cite{Avron}. The sign in the denominator comes from the dispersion relation of light which features both positive and negative branches. The Zeldovich formula and the "norm" used in \cite{Silke} writes as the ratio between a wave energy (which scales with the square of an amplitude) and the wave frequency. The case of acoustics is discussed in \cite{Stone} following the treatment by Landau and Lifschitz \cite{LL}.

The equivalence between the norm and the wave action density can be formally proven as follows.  First, wave packets on the free surface of water obey the Beltrami-Laplace equation in the Painlev\'e-Gullstrand metric as shown by Schutzhold and Unruh in 2002 (see \cite{SU} and the corresponding chapter in this book):

\begin{equation}
\partial_t(\partial_t\phi+U\partial_x\phi)+\partial_x(U\partial_t\phi+U^2\partial_x\phi)-c^2\partial_x^2\phi=0
\end{equation}
where $\phi$ is the velocity potential fluctuation. The complete velocity potential featuring both the waves and the background flow $U$ is such that its space derivative is by definition the flow velocity.

We can expect the conservation of two quantities due to the invariance of the corresponding action under 1)  the transformation $\phi\rightarrow e^{i\alpha}\phi$, $\alpha$ constant, and  2) time translation (for time-independent $U$). The former invariance gives conservation of the Klein-Gordon norm (as demonstrated elsewhere in this book):
\begin{equation} \label{norm}
N=\frac{i}{2c^2}\int_{-\infty}^\infty dx\left[\phi^*(\partial_t\phi+U\partial_x\phi)-\phi(\partial_t\phi^*+U\partial_x\phi^*)\right],
\end{equation}
whereas the latter gives conservation of (pseudo-)energy. For wave packets confined to a region where the flow velocity $U$ is constant, the norm (\ref{norm}) can be written in $k$-space in terms of the Fourier transform $\tilde{\phi}(k)$ as:
\begin{equation} 
\label{normk}
N=\frac{1}{c^2}\int_{-\infty}^\infty dk(\omega-Uk)|\tilde{\phi}(k)|^2,
\end{equation}

The typical interpretation of the Zeldovich formula is that it is a positive quantity: the number of photons. However, the Klein-Gordon norm used but the relativists is either positive or negative. Then, the Zeldovich formula encodes in general a different information than the Klein-Gordon norm. The latter counts the amount of charge, that is why for real fields it is zero. The complex solutions however do have charge. In fact, strictly speaking, the last equation is not correct as for a single k there can be modes with $\pm |k|$. Thus, apart from the integral there should be a sum in positive/negative branches:
\begin{equation} 
N= \frac{1}{ c^2}\int dk (c|k| |a_k|^2 - c|k| |b_k|^2).
\end{equation}
with $a_k$, $b_k$ the corresponding Fourier coefficients.

The Zeldovich formula (without the negative sign in the denominator) does not give zero for real fields. It really corresponds to$\frac{1}{c^2}\int dk (c|k| |a_k|^2)$. It does not contain the second term that in the case of real fields ($a_k=b_k$), would combine to yield a total zero (this is the case also for photons). The point is that what corresponds to the norm is not one of the pieces individually but the addition of the two.

In general, the norm scales like the integral over the wavenumber of the amplitude square of the Fourier transformed velocity potential  times the relative frequency in the moving frame. Hence, the norm scaling is $N \approx \int dk(\omega-Uk)\tilde{\phi}^2$ \cite{Silke}. However, because the velocity potential is related to the free surface deformation $\eta$ by the Bernoulli equation $\partial \phi/\partial t + g \eta =0$ (here, without a flow to simplify), it follows that the velocity potential scales like $\tilde{\phi} \approx g\tilde{\eta}/(\omega -Uk)$ in the Fourier space \cite{SU}. We conclude that the norm behaves like $N \approx \int dk g^2 \tilde{\eta}^2 /(\omega-Uk)$ as the wave action that is as the ratio between the square of the amplitude (the energy) and the relative frequency.

The norm is strictly conserved. Is this the case for the wave action ? In the fluid mechanics literature, the wave action is the solution of a conservation equation which replaces obviously the conservation of energy of a closed system. Here the system is open since the waves interact with the flow and do exchange energy. Bretherton and Garrett have shown that the wave action conservation writes in the so-called WKBJ regime where the flow velocity varies on a length scale much larger that the wavelength \cite{BG, Ivar}:
\begin{equation}
\frac{\partial }{\partial t} \left( \frac{E'}{\omega '} \right) + \nabla . \left(c_{g} \frac{E'}{\omega '} \right)=0
\end{equation}
where $\omega '=\omega -Uk$ and $c_g$ is the total group velocity including the background flow. According to Bretherton and Garrett, ``{\it because E' is an energy density, it is not constant down a ray, even if wave energy is conserved. However, in a time dependent and/or non-uniformly moving
medium, $\omega'$ varies along a ray. If $E'/\omega'$ is the wave action density, total wave action is
conserved, whereas total wave energy is not}". For a stationary process, we deduce that $\eta ^2 c_g/(\omega -kU)= \mathrm{const}$ since the energy density in the moving $E'$ is proportional to the square of the interface deformation $\eta$ (as in the rest frame without current). As a consequence, the amplitude diverges to infinity if one gets close to a turning point where the group velocity vanishes and where the WKBJ approximation is no longer valid. Dispersion enters the game to avoid such a caustic.

The change of wave action $J=E/\omega$ of a slowly modulated oscillator is exponentially small in the non-adiabatic parameter ($\omega$ / (rate of change of the medium properties)) : a mathematical theorem due to Meyer in 1973 \cite{Meyer}. For the linear pendulum of Rayleigh with a varying length, the rate of change is directly the inverse of the time lapse. Here, with water waves on a non-uniform flow, the property is the velocity $U$ and its typical rate of change is its space gradient (the so-called surface gravity in General Relativity) whose dimension is the one of a frequency: Jacobson and Parentani defined the surface gravity as a local expansion rate seen by a freely falling observer when he crosses the horizon \cite{JR}. One is tempted to extrapolate the following behavior for the change of wave action as the waves propagate against the flow:
\begin{equation}
\Delta J = J_0 \exp\left(-constant \frac{\omega}{\kappa}\right)
\end{equation}
where $\kappa = dU/dx$ is the surface gravity for water waves.

The change of wave action would be very similar to the famous Hawking spectrum \cite{BLV}: 
\begin{equation}
\frac{\beta ^2}{\alpha ^2} = \exp\left(-2\pi \frac{\omega}{\kappa}\right)
\end{equation}

The fact that the bogoliubov coefficient behaves as an exponential has been discussed by Jacobson \cite{CECS}.

Let us introduce the following dimensionless numbers $\omega$ / (rate of change of the medium properties) with names of distinguished physicists :
\begin{equation}
\frac{\omega }{\kappa} = \mathrm{Hawking} \quad \mathrm{number}=\mathscr{H}_w 
\end{equation}
\begin{equation}
\frac{\omega}{\frac{dU}{dx}} = \mathrm{Unruh} \quad \mathrm{number}=\mathscr{U}_n
\end{equation}



The validity of the WKBJ inequality $\frac{U}{\frac{dU}{dx}}>>\lambda$ can be reassessed thanks to the Unruh number close to the horizon. As a matter of fact, $U^* \approx g/\omega$ and $\lambda ^* \approx g/\omega^2$ then $\mathscr{U}_n \simeq O(1)$: close to the horizon, the WKBJ approximation breaks down. If $\mathscr{U}_n>>1$, then the process is adiabatic. In order not to have a vanishing spectrum, $\mathscr{U}_n \approx O(1)$, then vacuum radiation \`a la Hawking-Unruh is a non-adiabatic process \cite{Massar}. The case $\mathscr{U}_n<<1$ would imply a too small frequency, hence the amplitude of the energy spectrum (which scales with the cube of the frequency in 3D) would vanish.

\section*{Conclusion}

This rapid tour of the field of analogue gravity through the prism of water waves theory has broadened our definition of a horizon and has deepened our understanding of the concept of norm as used by relativists. In a related chapter of this book, we study experimentally the influence of surface tension and the associated dispersive horizons. Moreover, we try to answer to the question ``what is a particle close to a horizon? ".

\section*{Acknowlegments}

I would like to thank Thomas Philbin, Gil Jannes, Carlos Barcelo and Iacopo Carusotto for very interesting remarks which improve the content of this chapter.

\end{document}